\documentstyle [12pt]{article}
\newcommand{\be}{\begin{equation}}
\newcommand{\ee}{\end{equation}}
\newcommand{\bear}{\begin{eqnarray}}
\newcommand{\eear}{\end{eqnarray}}

\newcommand{\M}[2]{\mbox{$M_{#1}^{#2}$}}

\newcommand{\al}{\alpha}
\newcommand{\bt}{\beta}

\newcommand{\p}{\pi}

\newcommand{\omm}{\omega_{max}}

\newcommand{\orr}{\omega_{R}}

\newcommand{\as}{\alpha_{s}}

\begin{document}
\pagenumbering{arabic}
\addtocounter{page}{0}

\thispagestyle{empty}
\begin{flushright}
   \vbox{\baselineskip 12.5pt plus 1pt minus 1pt
         SLAC-PUB-6099 \\
         April 1993 \\
         (T/E)
             }
\end{flushright}

\begin{center}
{\bf Observation of High Energy Quark-Antiquark Elastic \\
  Scattering with Mesonic Exchange\footnote{Work supported
 in part by  Department of Energy
 contract DE-AC03-76SF00515.}}

Wai-Keung Tang \\
\vskip 1\baselineskip
Stanford Linear Accelerator Center \\
Stanford University, Stanford, CA 94309
\end{center}

\medskip


\begin{abstract}
 We studied the high energy quark anti-quark elastic scattering
  with an exchange
of a mesonic state in the $t$ channel with $-t/\Lambda_{QCD}^{2}
\gg 1$. Two  methods are proposed to eliminate the strong
background from bare pomeron, reggized gluon and odderon exchange. The
feasibility of measuring mesonic reggeon exchange is discussed.
\end{abstract}

\begin{center}
(Submitted to \it{Physics Letters} \bf{B})
\end{center}
\newpage
\noindent{\bf Introduction}

With the advances of Tevatron, LHC and SSC, it is possible to study
experimentally
the Regge behavior in the parton level where the momentum transfer
squared $-t \gg \Lambda_{QCD}^{2}$ but is still smaller than the center of mass
energy squared $\hat{s}$
of the partons.  The presence of a ``large" scale $-t$ justifies
 the use of perturbative QCD.
The dominant process is an exchange of a
Balitsky-Fadin-Kuraev-Lipatov
(BFKL) pomeron~[1,4] between partons.  Several hard partonic processes  to
measure the behavior
of the BFKL pomeron [5-10] have been discussed in the
literature.
  In Ref \cite{tam},
A.H.~Mueller and the author propose a process of high energy,
 fixed $t$ parton-parton elastic scattering through the exchange of
a BFKL pomeron.  The elastic scattering produces a final state which,
at the parton level, contains two jets with a rapidity gap
i.e. a region of rapidity in which no gluons are found.
It is natural to extend this idea to the
process whereby a mesonic reggeon is exchanged \cite{wkt}. Even though
the mesonic
exchange process is suppressed by a factor of $\hat{s}^{2}$ with respect to
pomeron exchange, the reggeon trajectory is itself interesting and it
opens another window to test PQCD in  double logarithmic
approximation.

The energy dependence of the scattering cross section by  exchanging
 a mesonic reggeon is $\hat{s}^{-2+2\omega_{R}}$ where
 $\omega_{R}$, $ \sim 0.2$, in the interesting physical region \cite{wkt}, is
the trajectory of
 the
  mesonic reggeon. The smalless of the trajectory and the suppression
  factor $-2$ in the energy dependence impose a serious obstacle to
  observing mesonic reggeon behavior. Thus, the rapidity  between the
  two partons cannot be taken to be too large,
otherwise the cross section will be too small to be observed.

At the parton level, there are several potential backgrounds to the signal of
observing the interesting
rapidity gap phenomena by exchanging a mesonic reggeon. The dominant
one is the BFKL pomeron exchange, as mentioned above.
Reggized gluon exchange also
contributes if we assume that we cannot detect partons with transverse
momentum smaller than  a fixed parameter $\mu$.
It can be
of the same order as pomeron exchange if rapidity  is not too large
\cite{tangdelduca}.  Thus, background from reggized gluon can be
substantial. Even though quantitative prediction for the exchange of
perturbative odderon is still lacking, we expect that it will not be too
small,
and may be comparable with that of  mesonic reggeon exchange \cite{Mueller}.
Thus, two methods to eliminate the strong background due to
the exchange of pomeron, reggized gluon and odderon are suggested below.

        The simplest way to
eliminate all these background is to do $A\bar{B}$ and $AB$
collisions with two jets, tagged  to be  $q_{i}\bar{q}_{i}$ pairs, but with
$q_{i}\bar{q}_{i}$
not the constituent quarks
of hadrons $A,\;B,$ and $\bar{B}$, e.g. $s\bar{s},\;c\bar{c},$ and $d\bar{d}$
in
$P\bar{P}$ and $PP$ collisions.
Obviously, pomeron and reggized gluon contribution are subtracted out if we
take the difference of the cross sections. The elimination of the
background due to odderon through the tagging of apppropiate $q_{i}\bar{q}_{i}$
pairs
needs some explanation. In general, the amplitudes of
scattering of $q_{i}q_{j}$ and $q_{i}\bar{q}_{j}$ through the exchange
of an odderon have a phase
difference  $\pi/2$. If we consider the difference the cross sections
 of two jets
 of $A\bar{B}$ and $AB$ collisions, odderon contribution cannot
be eliminated. Instead, cross sections of
$q_{i/A}\;\bar{q}_{j/\bar{B}} \rightarrow q_{i}\bar{q}_{j}$ and
$q_{i/A}\;q_{j/B}
\rightarrow q_{i}q_{j}$
scattering, where $q_{i/A}$ stands for quark $q_{i}$ from hadron $A$,
  add  due to the $\pi/2$ phase difference  in their
amplitudes.
However, the tagging of  the final jets, taking $P\bar{P}$ and $PP$
collisions as an example, being $s\bar{s},\;c\bar{c},\;d\bar{d}$, eliminates
odderon contributions completely, since there is complete cancellation
between the scattering of $q_{i/P}\;\bar{q}_{i/\bar{P}} \rightarrow
q_{i}\bar{q}_{i}$ and
$q_{i/P}\;\bar{q}_{i/P} \rightarrow q_{i}\bar{q}_{i}$ if one assumes
parton distribution of
$\bar{q}_{i/\bar{P}}$ and $\bar{q}_{i/P}$ are the same for
$\bar{q}_{i}=\bar{s},\;\bar{c},\;\bar{d}$.

The
tagging of quarks in the extremely forward direction may have a
serious efficiency problem,
so we also suggest another method which
exploits the fact that mesonic reggeon and pomeron (reggized gluon
and
odderon) couple
differently to quark (anti-quarks) with different helicity states.
A mesonic reggeon only couples to an initial  quark anti-quark pair when they
have opposite
 helicities. Pomeron, reggized gluon and odderon are insensitive to the
helicities of quark and
anti-quark. Thus, the difference of the cross sections of
$A_{\uparrow}B_{\uparrow}$ and
$A_{\uparrow}B_{\downarrow}$ collisions eliminates the contribution from
pomeron, reggized gluon and odderon completely and only indicates the behavior
of mesonic exchange. Chosing
$B$ to be the anti-particle of $A$ would maximize the cross section
since parton distribution of $\bar{q}_{i}$
in $B=\bar{A}$ is of the same order as parton distribution of $q_{i}$
in $A$. However, luminousity of an $A\bar{A}$ machine is limited due to the
difficulty of
producing a large number of anti-particles. It is not yet clear  what is
the best way to observe mesonic exchange phenomena.

Unfortunately, the above two methods
also eliminate the contribution due to the quark quark
backward scattering with
mesonic exchange. It is obvious that cross section of quark quark
scattering cancel completely once the difference of the cross sections of
$ P\bar{P}$
and $PP$ collision is taken.  As  mesonic reggeon can couple to an
initial quark quark pair with any helicity
states, quark quark backward scattering cannot contribute in the
second method. Only the
quark antiquark
elastic scattering with   mesonic reggeon exchange survives.

At the hadron level, the spectator interactions
radiate soft gluons, producing hadrons across the  rapidity interval
and thus spoiling the rapidity gap. This soft
gluon radiation is essential to prove the QCD factorization theorem, which
relates parton to hadron cross section \cite{jccollin}. Thus, we have
two options \cite{tangdelduca}: ($i$) we require that the soft gluon
radiation is suppressed and estimates the rapidity-gap survival
probability \cite{bj}: ($ii$) we allow the presence of soft gluons with
transverse momentum $\mu \gg \Lambda_{QCD}$ \cite{tam}. Then QCD factorization
holds, and one can use it to relate parton to hadron cross sections. In
Ref.
\cite{tangdelduca}, this
process is named quasi elastic scattering. In this paper we take the
second approach.

\noindent{\bf Quark Anti-quark Elastic Scattering}

We study the quark anti-quark scattering
amplitudes with mesonic exchange in the $t$ channel. In the
kinematic region,
\be
\hat{s} \simeq \mid \hat{u} \mid \gg -t  \gg \Lambda_{QCD}^{2},  \nonumber
\ee
 higher order corrections  like
$\sim \al_{s} [(\al_{s}/\p) y^{2}]^{n}$ are important, when $ y$, the
rapidity between a quark and an anti-quark, defined by $y=\ln(\hat{s}/-t)$, is
large.
The strong coupling constant  $\alpha_{s}$ is evaluated at $-t$.
This is the double logarithmic (DL) approximation. The method of
separating the softest virtual particle \cite{ki} allows one to calculate the
partial wave amplitudes in the double logarithmic approximation
\cite{li}. The spinor structure of the Born
term is preserved in higher order, so the scattering amplitudes of
$q_{i}\bar{q}_{i}\rightarrow q_{j}\bar{q}_{j}$ through color singlet
(the color octet is suppressed so we will not consider it) mesonic exchange
can be written as \cite{li}
\be
A^{p}(\hat{s},t) =  \frac{\gamma_{\mu}^{\perp}
\otimes
\gamma_{\mu}^{\perp}}{\hat{s}}\frac{1}{N_{c}}\delta_{aa'}\delta_{bb'}
\M{}{p}(y),
\ee
where $a,b$ and $a',b'$ label the color states of the initial
and final quarks and anti-quarks.  The signature
corresponding to even (odd) parts of the amplitude under the
transformation $\hat{s}\leftrightarrow \hat{u} \simeq -\hat{s}$ is
$p=1\;\;(-1)$.
$\M{}{p}(y)$
gives all order corrections. Partial wave amplitude is
evaluated by R.~Kirschner and L.N.~Lipatov \cite{li}, and the expression
of the ampltiude can be obtained by performing a Mellin
transformation \cite{wkt}. As shown in Ref.\cite{wkt},
the difference between the magnitude of positive and negative signature
singlet
amplitudes with fixed strong coupling can be practically neglected up to SSC
energies.
They only have a phase difference  of $\pi/2$.
However, fixed coupling is not a good approximation, as the transverse
momentum in the loop extends over the large range from $-t$ to
$\hat{s}$. It is expected  that the running coupling effect changes the
amplitude greatly. We observed that behavior. The nature of the
partial wave
singularity  changes from a square root branch cut to infinitely many poles
\cite{wkt2}. However, inclusion of running coupling decreases the
difference between  positive and negative signature singlet
amplitudes for the following reasons: the only
difference between the positive and negative channels stems from
the contribution of the double logarithmic soft gluon;
the soft gluon
 contributes to the negative signature singlet channel but not
the positive signature singlet channel~\cite{li}.
  The inclusion of the running coupling
will decrease the importance of the soft gluon contribution, as the
coupling between quark and gluon is smaller than in the
fixed coupling case. Therefore, for practical purposes, we can
 take positive and negative singlet amplitudes
 to be equal in magnitude but with phase difference $\pi/2$. The
 positive signature amplitude with running coupling constant is
 \cite{wkt}
 \bear
\M{run}{+}(y) &=& 0.895\times 64\p^{3} b \, \orr^{2}
                                 (\frac{\as}{2\p a_{0}})^{1/2}
                  e^{\orr y},
\eear
with
\be
a_{0}=\frac{N_{c}^{2}-1}{2N_{c}};\;\;\;\;
b=\frac{1}{16\pi^{2}}(\frac{11}{3}N_{c}-\frac{2}{3}N_{f}),
\ee
where $N_{c}$ $(N_{f})$ is the number of colors (flavors) and $\orr(\as)$ is
the
leading trajectory of the mesonic reggeon. It is a
non-linear function of $\alpha_{s}(-t)$.
Let $\omega_{max}=a_{0}/(8\pi^{2}b)$
and the leading mesonic
trajectory can be written as
\bear
\orr(\as) &=& \frac{\omm}{\rho_{1} + \rho_{2} \as^{-1/2}}
\label{tra}
\eear
with
\bear
\rho_{1} =1.14 \times \frac{3}{4} \;\;\;\;\;& & \rho_{2}=0.90 \times
         \frac{2}{\pi}(\frac{\omm}{4\pi b})^{1/2}.
\eear
The numerical factors $0.895$ in the normalization and $1.14$ and $0.90$
in the trajectory are the result of numerical fitting. The trajectory
is of the order  $0.2$ in the interesting physical range. It's
smallness imposes serious difficulty in observing the mesonic reggeon
trajectory in the parton level experimentally. Sandwiching
$A^{p}$ between appropiate spinors and averaging (summing) over initial
(final) color states, we have, when the quark and anti-quark have opposite
helicities,
\be
|A^{+}_{run} (y)|^{2} = \frac{4}{N_{c}^{2}} |\M{run}{+}(y)|^{2} .
\ee
Here, we assume that the mass of quarks can be neglected when compared
with the parton-parton center of mass energy $\sqrt{\hat{s}}$. Therefore,
the differential cross section of $q_{i}\bar{q}_{i} \rightarrow
q_{j}\bar{q}_{j}$ is
\bear
\frac{d\hat{\sigma}_{\al\bt}}{dt} &=& \frac{1}{16\pi\hat{s}^{2}}
|A^{+}+A^{-}|^{2} \delta_{\al,-\bt} \nonumber \\
&\simeq& \frac{1}{2\pi N_{c}^{2}\hat{s}^{2}}
|\M{run}{+}|^{2} \delta_{\al,-\bt}
\label{hardcrosssection}
\eear
where $A^{-} \simeq iA^{+}$ and $\al (\bt)$ is the helicity of the initial
quark
(anti-quark).

\noindent{\bf Observation of Mesonic Reggeon Exchange}

{}From Ref. \cite{tam}, the most straightforward way to
observe the mesonic scattering in the parton level is to look at the
cross section of two jets, $x_{A}x_{B}d\sigma/dx_{A}dx_{B}dt$
 where $x_{A}$ and $x_{B}$ are the
longitudinal momenta fractions with respect to their parent hadrons.
The two tagging jets must have nearly balancing transverse
momenta and that no additional jets, above a transvese momentum scale
$\mu$, be found in the rapidity interval between the tagging jets.
Then, cross section due to
quark anti-quark elastic scattering with mesonic reggeon exchange is
\bear
\!\!\!\!\!\!\!\!\!\!\!& &x_{A}x_{B}\frac{d\sigma}{dx_{A}dx_{B}dt}(A_{\lambda}
B_{\lambda'} \rightarrow
j(x_{A})j(x_{B})) \hfill\nonumber \\
\!\!\!\!\!\!\!\!\!\!\!&= & N\, \sum_{f}
\;\left[\;x_{A}x_{B}\,Q_{f/A}^{\al/\lambda}(x_{A})\,
\bar{Q}_{f/B}^{\bt/\lambda'}(x_{B}) + (Q \leftrightarrow \bar{Q})\; \right]
\frac{d\hat{\sigma}_{\al\bt}}{dt}
\eear
where $Q_{f/A}^{\al/\lambda}$ $(\bar{Q}_{f/A}^{\al/\lambda})$ is the quark
(anti-quark) distribution with
helicity $\al$ in hadron $A$ with polarization $\lambda$. $N$ is the
number of flavors of quark anti-quark pairs allowed in the final jets.
For $P\bar{P}$ and $PP$ collisions $N=3$, and the final quark anti-quark pair
should be
$s\bar{s},\;c\bar{c},\;d\bar{d}$  so that background from pomeron,
reggized gluon and odderon can be eliminated. The differential cross
section for this process is
\bear & &
x_{A}x_{B}t\left[\frac{d\sigma(P\bar{P})}{dx_{A}dx_{B}dt}-
\frac{d\sigma(PP)}{dx_{A}dx_{B}dt}\right]
\hfill
\nonumber \\
&=&
 \frac{N}{2}\,\sum_{f}\left[
\;x_{A}x_{B}\, Q_{f/P}(x_{A})\,
\Delta\bar{Q}_{f/P}(x_{B}) +(Q
\leftrightarrow \bar{Q}) \;\right]t\frac{d\hat{\sigma}}{dt}
\eear
with
\bear
\Delta Q_{f/P} &=& Q_{f/\bar{P}}-Q_{f/P} \nonumber \\
\Delta\bar{Q}_{f/P} &=& \bar{Q}_{f/\bar{P}}-\bar{Q}_{f/P}
\eear
The hard
scattering cross section
$d\hat{\sigma}/dt$ is the usual one defined by Eq.\ref{hardcrosssection}
without the delta function $\delta_{\al,-\bt}$.
We are also interested in the difference of $d\sigma(A_{\uparrow}B_{\uparrow})$
and
$d\sigma(A_{\uparrow}B_{\downarrow})$. In this case, $N=5$ and one can write,
\bear & &
x_{A}x_{B}t\left[\frac{d\sigma(A_{\uparrow} B_{\downarrow} )}{dx_{A}dx_{B}dt}-
\frac{d\sigma(A_{\uparrow}
B_{\uparrow})}{dx_{A}dx_{B}dt}\right]
\hfill
\nonumber \\
&=&
 N\,\sum_{f}\left[
\;x_{A}x_{B}\,\delta Q_{f/A}(x_{A})\,
\delta\bar{Q}_{f/B}(x_{B}) +(Q
\leftrightarrow \bar{Q}) \;\right]t\frac{d\hat{\sigma}}{dt}
\eear
with
\bear
\delta Q_{f/A}&=&Q_{f/A}^{\uparrow}-Q_{f/A}^{\downarrow} \nonumber \\
\delta \bar{Q}_{f/A}&=&\bar{Q}_{f/A}^{\uparrow}-\bar{Q}_{f/A}^{\downarrow}
\eear
where $\uparrow$ $(\downarrow)$ in the superscript of $Q$ means the helicity
of the quark is
parallel (anti-parallel) to that of the parent hadron.

The hard scattering cross
section has a suppression factor $\hat{s}^{-2}$ so we want to minimize
$\hat{s}$ with a reasonably large rapidity gap. Taking $y=4$ and $-t=25
\mbox{GeV}^{2}$; $\hat{s} \simeq 1365 \mbox{GeV}^{2}$,
$\alpha_{s}=0.223$, $\omega_{R}=0.189$ and $td\hat{\sigma}/dt \sim 120
\mbox{pb}$, if $x_{A}=x_{B}=0.1$, the difference of the cross sections of
$P\bar{P}$ and $PP$ collisions written in Eq.10 is $ \sim
90\mbox{pb}$ which is not too small and can be measured. If we take
$A=P$ and $B=\bar{P}$, Eq.12 should give us the same order of
magnitude but with the advantage of not tagging the final jets. In
sum, the observation of mesonic reggeon exchange is possible, and it
will be very interesting to measure the trajectroy of the mesonic
reggeon.

\begin{flushleft}{\bf Acknowledgements}\end{flushleft}

I am very much indebted to A.H.Mueller for
suggesting  this work and for many stimulating discussions.
Also thanks to Stanley Brodsky for continous encouragement and
discussions.

\end{document}